\documentclass[reqno]{article}
\newcommand{\ds}{\displaystyle}
\newcommand{\dsf}{\ds\frac}

\newcommand{\beq}{\begin{equation}}
\newcommand{\eeq}{\end{equation}}

\usepackage{ae} % or {zefonts}
\usepackage[T1]{fontenc}
\usepackage[ansinew]{inputenc}
\usepackage{amsmath}
\usepackage{graphicx}
\usepackage{color}
\usepackage[colorlinks]{hyperref}
\usepackage{multicol}
\begin{document}
\footnotesize
\begin{center}
\bf Blow-up instability in the vortex sate of type II
superconductors
\end{center}

\begin{center}
N. A. Taylanov and A. S. Rakhmatov
\end{center}

\begin{center}
\emph{National University of Uzbekistan}
\end{center}

\begin{center}
{\bf Abstract}
\end{center}
\begin{center}
\mbox{\parbox{13cm}{\footnotesize The spatial and temporal
evolution of small perturbations of the temperature and
electromagnetic field for the simplest geometry, i.e., in a
superconducting slab placed in a parallel magnetic field is
considered.  The obtained solution describes a blow-up-type
instability in the sample. It remains localized within the limited
area $x<L^*/2$ with increasing infinitively of time. }}
\end{center}
{\bf Key words}: nonlinear equations, thermal and electromagnetic
perturbations, critical state, flux creep.

\begin{multicols}{2}{The magnetic flux penetration into a type-II superconductor occurs
in the form of quantized vortices. In the presence of different
types of defects or pinning centers in the superconductor sample,
the vortices may be attached to such defects. A nature of
interaction between the vortices and the structural defects is
determined by the pinning force $F_P$. If transport current with
the density $j$ is passed through superconductor, the interaction
of the current with vortex lines leads to the emergence of the
Lorentz force $F_L$, acting on each one of the vortices. Under the
effect of the Lorentz force $F_L $ the viscous flux flow of
vortices begin to move. The viscous magnetic flux flow in
accordance with electromagnetic induction creates a vortex
electric field $E$. This means that energy dissipation occurs, an
electric resistance appears and the superconductor undergoes a
transition to the resistive or to the normal state. Propagating
magnetic flux causes Joule heating, giving rise to global and/or
micro flux avalanches in the critical state of type-II
superconductors. Thus, flux jumps results in a large-scale flux
avalanches in a superconductor and their origin are related to the
magnetothermal instabilities [1-5]. Thermomagnetic instability or
flux jump phenomena have been observed in conventional hard
superconductors [1-6], as well as in high-temperature
superconductors, recently [7, 8]. The critical state stability
against flux jumps in hard and composite superconductors has been
discussed in a number of theoretical and experimental papers
[1-5]. The general concept of the thermomagnetic instabilities in
type-II superconductors was developed in literature [4, 5]. The
dynamics of small thermal and electromagnetic perturbations, whose
development leads to the flux jump, have been investigated
theoretically in detail by Mints and Rakhmanov [5]. The authors
have found the stability criterion for the flux jumps in the
framework of adiabatic and dynamic approximations in the viscous
flux flow regime of type-II superconductors. Conventionally,
thermomagnetic instabilities were interpreted in terms of thermal
runaway triggered by local energy dissipation in the sample [5].
According to this theory, any local instability causes a small
temperature rise, the critical current is decreased and magnetic
flux moves much easily under the Lorentz force. The additional
flux movement dissipates more energy further increasing
temperature. This positive feedback loop may lead to a flux jumps
in the superconductor sample.

In our previous work, the dynamics of small thermal and
electromagnetic perturbations has been studied in the flux flow
regime, where voltage current-current characteristics of hard
superconductor is described by linear dependence of $j(E)$ at
sufficiently large values of electric field [9]. In the region of
weak electric fields the current-voltage characteristics $j(E)$ of
superconductors is highly nonlinear due to thermally activated
dissipative flux motion.  A theoretical analyze of the flux
jumping in the flux creep regime, where the current-voltage
characteristics of a sample is a nonlinear have been carried out
recently by Mints [10] and by Mints and Brandt [11]. However, a
careful study the dynamics of the thermal and electromagnetic
perturbations in the regime weak electric field with nonlinear
current-voltage characteristics associated with flux creep is
still lacking.

\vskip 0.5 cm In the present paper the spatial and temporal
evolution of small perturbations of the temperature and
electromagnetic field in the flux creep regime for the simplest
geometry, i.e., in a superconducting slab placed in a parallel
magnetic field is considered. We shall study the problem in the
framework of a macroscopic approach, in which all lengths are
larger than the flux-line spacing; thus, the superconductor is
considered as a uniform medium.

   We study the evolution of the thermal and electromagnetic
penetration process in a simple geometry - superconducting
semi-infinitive sample $x\geq 0$. We assume that the external
magnetic field induction $B_e$ is parallel to the z-axis and the
magnetic field sweep rate $\dot{B_e}$ is constant. When the
magnetic field with the flux density $B_e$ is applied in the
direction of the z-axis, the transport current $\delta j(x, t)$
and the electric field $\delta E(x, t)$ are induced inside the
slab along the y-axis. For this geometry, the temporal and spatial
evolution of thermal $\delta T(x, t)$, electromagnetic field
$\delta E(x, t)$ and current $\delta j(x, t)$ perturbations are
described by the following nonlinear heat diffusion equation
[9-11]

\begin{equation}
\nu\dsf{d\delta T}{dt}=k\dsf{d^2\delta T}{dx^2}+j_c\delta E.
\end{equation}
\begin{equation}
\dsf{d^2\delta E}{dx^2}=\mu_0\left[\dsf{dj}{dE}\dsf{d\delta
E}{dt}-\dsf{dj_c}{dT}\dsf{d\delta T}{dt}\right].
\end{equation}
Here $j_c$ is the critical current density, $\nu=\nu(T)$ and
$\kappa=\kappa(T)$ are the specific heat and thermal conductivity,
respectively. In order to obtain analytical results of a set Eqs.
(1), (2), we suggest that $j_c$ is independent on magnetic field
induction $B$ and use the Bean's [1] critical state model
$j_c=j_c(B_e, T)$

$$
j_c(T)=j_c(T_0)\left[1-\dsf{T-T_0}{T_c-T_0}\right],
$$
where $j_c(T_0)$ is the equilibrium current density, $T_0$ and
$T_c$ are the equilibrium and critical temperatures of the sample,
respectively [9]. For the sake of simplifying of the calculations,
we perform our calculations on the assumption of negligibly small
heating $(T-T_0\ll T_c-T_0)$ and assume that the temperature
profile is a constant within the across sample and thermal
conductivity $\kappa$ and heat capacity $\nu$ are independent on
the temperature profile.  We shall study the problem in the
framework of a macroscopic approach, in which all lengths scales
are larger than the flux-line spacing; thus, the superconductor is
considered as a uniform medium.

The set of differential equations (1), (2)) should be supplemented
by a current-voltage curve $j=j(E, B, T)$. In the flux creep
regime the current-voltage characteristics of type - II
conventional superconductors is highly nonlinear due to thermally
activated dissipative flux motion [12, 13]. For the logarithmic
current dependence of the potential barrier $U(j)$, proposed by
[14] the dependence $j(E)$ has the form

\begin{equation}
\vec j=j_c\left[\dsf{\vec E}{E_c}\right]^{1/n},
\end{equation}
where $E_c=const$ and the parameter n depends on the pinning
regimes and can vary widely for various types of superconductors.
In the case $n=1$ the power-law relation (3) reduces to Ohm's law,
describing the normal or flux-flow regime [6]. For infinitely
large $n$, the equation describes the Bean critical state model
$j\simeq j_c$ [1]. When $1<n<\infty$, the equation (3) describes
nonlinear flux creep [15]. In this case the differential
conductivity $\sigma$ is determined by the following expression

\begin{equation}
\sigma=\dsf{dj}{dE}\approx \dsf{j_c}{nE}.
\end{equation}
It is assumed, for simplicity, that the value of n temperature and
magnetic-field independent. It should be noted that the nonlinear
diffusion-type equations (1) and (2), completed by the flux creep
equation (4), totally determine the problem of the space-time
distribution of the temperature and electromagnetic field profiles
in the flux creep regime with a nonlinear current-voltage
characteristics (3) in a semi-infinite superconductor sample.

To find an analytical solution of Eqs. (1) and (2) we use simple
adiabatic approximation, assuming that $\tau\ll 1$, i.e., that the
magnetic flux diffusion is faster than the heat flux diffusion [4,
5]. Therefore, we neglect the diffusive term in the heat equation.
Then eliminating the variable $\delta T(x, t)$ by using the
relationship (1) and substituting into Eq. (2), we obtain a
second-order differential equation for the distribution of small
electromagnetic perturbation $e(x, t)$ in the form

\begin{equation}
\dsf{d^2e}{dz^2}=e^{\gamma}\dsf{de}{d\tau}-\beta e.
\end{equation}

Here, we introduced the following dimensionless variables
$$
z=\dsf{x}{L},\quad \tau=\dsf{t}{t_0},\quad \epsilon=\dsf{\delta
E}{E_c};
$$
and parameters
$$
\beta=\dsf{\mu_0 j_{c}^{2}L^2}{\nu(T_c-T_0)},\quad
\gamma=\dsf{1-n}{n},\quad t_0=\dsf{\mu_0 j_cL^2}{E_c}.
$$
Here $L=\dsf{cH_e}{4\pi j_c}$ is the magnetic field penetration
depth. Since we have neglected the redistribution of heat in
deriving Eq. (5), only the electrodynamic boundary conditions
(see, Ref. [9]) should be imposed on this equation

\begin{equation}
e(1, \tau)=0, \qquad \dsf{de(0, \tau)}{dz}=0.
\end{equation}

The solution of (5) together with the boundary conditions (6) can
be obtained by using the method of separation of variables.
Looking for the solution of Eq. (5) in the form

\begin{equation}
e(z, \tau)=\lambda(\tau)\psi(z).
\end{equation}

we get the following expressions for a new variables
\begin{equation}
\dsf{d\lambda}{d\tau}=-k\lambda^{1-\gamma},
\end{equation}
\begin{equation}
\dsf{d^2\phi}{dz^2}=k\phi^{\gamma+1}-\beta\phi.
\end{equation}
By integrating equation (8) we easily obtain
\begin{equation}
\lambda(\tau)=(\tau_p-\tau)^{1/\gamma},
\end{equation}
where $\tau_p$ is the constant parameter, describing the
characteristic time of magnetic flux penetration profile;
$k=1/\gamma$. Now, integrating twice, the ordinary differential
equation for the function $\phi(z)$ with the boundary conditions
(6) and taking into account (10), we find the following an
explicit solution of (5) in the form [16]
\begin{equation}
e(z,\tau)=\left[\dsf{B}{\tau_p-\tau}\cos^2\dsf{2\pi}{L^*}z\right]^{n/n-1},
\end{equation}
$$
B=\dsf{ n^2}{1-n^2}\dsf{\beta}{2}, \quad
\dsf{1}{L^*}=\dsf{1-n}{4\pi n}\sqrt{\beta},
$$
The obtained solution (11) describes the distribution of the
electromagnetic field in the flux creep regime with a power-law
current-voltage characteristics. The solution describes an
blow-up-type instability in the superconductor sample. As easily
can be seen that the solution remains localized within the limited
area $x<L^*/2$ with increasing infinitively of time. This
characteristic phenomenon often occurring in nonlinear parabolic
problems is blow-up of solutions in finite time [16]. In other
words, the growth of the solution, becomes infinite at a finite
time $\tau_p$. Typical distributions of the electric field e(z,t)
determined from analytical solution (11) are shown in Figures 1
and 2 for the values of parameters $\tau_p$=1, n=3, n=11
$\beta\sim$ 0.5 and $L^*\sim$ 1. The time evolution of the
electric field is shown in Fig.3.

\begin{center}
\bf Conclusion
\end{center}

In conclusion, we have performed a theoretical study of dynamics
of small thermal and electromagnetic perturbations in type-II
superconductors in the flux creep regime in the framework
adiabatic approximation. For this purpose, the space-time
evolution of temperature and electric field was calculated using
the heat diffusion equation, coupled with Maxwell’s equations and
material law, assuming that heat diffusion is small that the
magnetic diffusion. An explicit solution of the diffusion equation
has been obtained, which describes the distribution of the
electromagnetic field in the flux creep regime with a power-law
current-voltage characteristics.

\begin{center}
\bf Acknowledgements
\end{center}

This study was supported by the NATO Reintegration Fellowship
Grant and Volkswagen Foundation Grant. Part of the computational
work herein was carried on in the Condensed Matter Physics at the
Abdus Salam International Centre for Theoretical Physics.

\begin{center}
\bf References
\end{center}

\begin{enumerate}

\item C. P. Bean, Phys. Rev. Lett. 8, 250, 1962; Rev. Mod. Phys.,
36, 31, 1964.

\item P. S. Swartz and S. P. Bean, J. Appl. Phys., 39, 4991, 1968.

\item S. L. Wipf, Cryogenics, 31, 936, 1961.

\item  R. G. Mints and A. L. Rakhmanov, Rev. Mod. Phys., 53, 551,
1981.

\item R. G. Mints and A. L. Rakhmanov, Instabilities in
superconductors, Moscow, Nauka, 362, 1984.

\item  A. M. Campbell and J. E. Evetts, Critical Currents in
Superconductors (Taylor and Francis, London, 1972)  Moscow, 1975.

\item L. Legrand, I. Rosenman, Ch. Simon, and G. Collin, Physica
C, 211, 239, 1993.

\item   A. Nabialek, M. Niewdzas, H. Dabkowski, J. P. Castellan,
and B. D. Gaulin, Phys. Rev., B 67, 024518, 2003.

\item N. A. Taylanov and A. Elmuradov, Technical Physics, 11, 48,
2003.

\item R. G. Mints, Phys. Rev., B 53, 12311, 1996.

\item R. G. Mints and E. H. Brandt, Phys. Rev., B 54, 12421, 1996.

\item P. W. Anderson , Y.B. Kim  Rev. Mod. Phys., 36. 1964.

\item P. W. Anderson,  Phys. Rev. Lett.,  309, 317, 1962.

\item E. Zeldov, N. M. Amer, G. Koren, A. Gupta, R. J. Gambino,
and M. W. McElfresh, Phys. Rev. Lett., 62, 3093, 1989.

\item P. H. Kes, J. Aarts, J. van der Berg, C.J. van der Beek, and
J.A. Mydosh, Supercond. Sci. Technol., 1, 242, 1989.

\item A. A. Samarskii, V. A. Galaktionov, S. P. Kurdjumov, and A.
S. Stepanenko, Peaking Regimes for Quasilinear Parabolic
Equations, Nauka, Moskow, 1987.

\end{enumerate}

}\end{multicols}

\newpage

\begin{center}
\includegraphics[width=3.5583in]{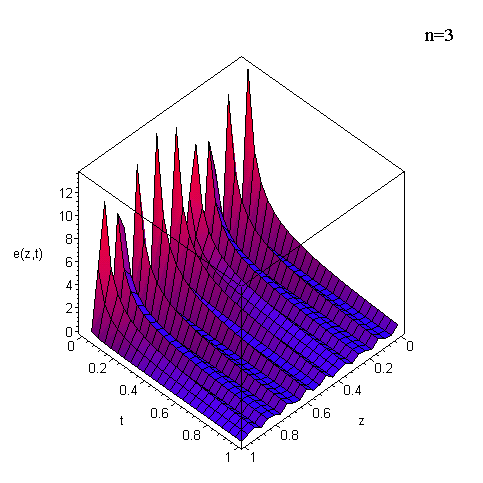}
\includegraphics[width=3.5583in]{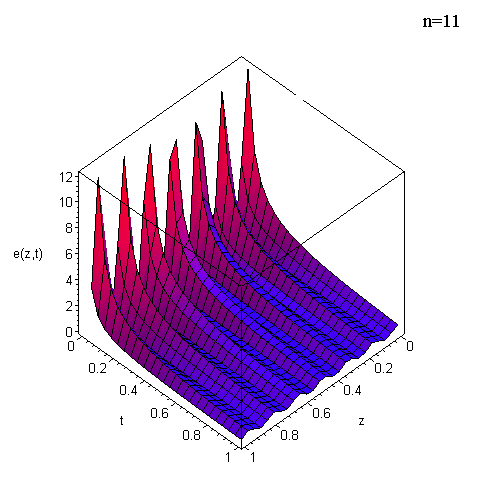}
\includegraphics[width=3.5583in]{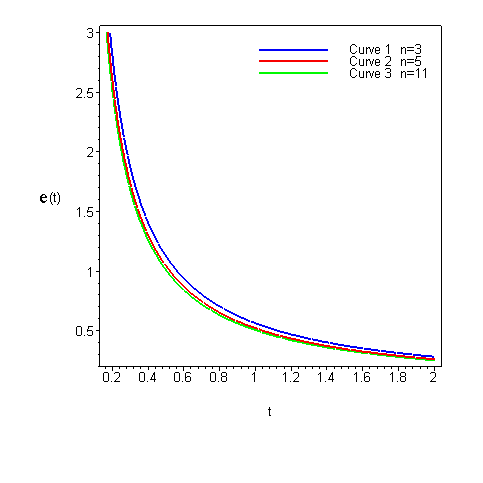}\\
\end{center}
\begin{center}
Fig.1 and 2. The space and time distributions of the electric
field profiles at different values of n=3 and n=11 at $\tau_p$=1,
$\beta\sim$ 0.5 and  $L^*\sim$ 1.

Fig.3. The time evolution of the electric field profile at
different values of n=3, n=7 and n=11 at $\tau_p$=1, $\beta\sim$
0.5 and  $L^*\sim$ 1.
\end{center}

\end{document}